# Low-energy band structure and even-odd layer number effect in AB-stacked multilayer graphene

Ryuta Yagi[1]*, Taiki Hirahara[1], Ryoya Ebisuoka[1], Tomoaki Nakasuga[1], Shingo Tajima[1], Kenji Watanabe[2], and Takashi Taniguchi[2]

[1] Graduate School of Advanced Sciences of Matter, Hiroshima University, Kagamiyama Higashihiroshima, Hiroshima 739-8530, Japan.
[2] National Institute for Material Science (NIMS), Namiki, Tsukuba 305-0044, Japan.

## Abstract

How atoms acquire three-dimensional bulk character is one of the fundamental questions in materials science. Before addressing this question, how atomic layers become a bulk crystal might give a hint to the answer. While atomically thin films have been studied in a limited range of materials, a recent discovery showing how to mechanically exfoliate bulk crystals has opened up the field to study the atomic layers of various materials. Here, we show systematic variation in the band structure of high mobility graphene with one to seven layers by measuring the quantum oscillation of magnetoresistance. The Landau fan diagram showed distinct structures that reflected differences in the band structure, as if they were finger prints of multilayer graphene. In particular, an even-odd layer number effect was clearly observed, with the number of bands increasing by one for every two layers and a Dirac cone observed only for an odd number of layers. The electronic structure is significantly influenced by the potential energy arising from carrier screening associated with a gate electric field.

*Corresponding author: yagi@hiroshima-u.ac.jp

## Introduction

Since the discovery of graphene [1], various theoretical and experimental studies have been carried out to elucidate its nature [2-9]. Moreover development of techniques to produce high-quality graphene [10-13] with sufficiently long mean free paths has made it possible to study ballistic transport [14-16] and to modify band structures by using moiré structures created by the slight misalignment of bilayer honeycomb lattices of graphene and $h$-BN [17]. Since the beginning of research on graphene, the electronic band structure of this material has been expected to show a particular layer number dependence. [18-24] Even if we consider only AB-stacking, the band structure should show an even-odd layer number effect [20-22, 24], as shown in Figure 1 **a**: the band structures of $2N$ and $2N + 1$ ($N$=1, 2, 3,..) layer graphene show $N$ bilayer-like bands with non-zero band mass, and the $2N + 1$ layer additionally shows a monolayer-like band that is massless with a linear dispersion relation. To date, optical measurements of the band structure have been performed on graphene consisting of several numbers of layers [25-28]. However, the low-energy band structure, which affects transport properties, has been only studied for samples consisting of a few layers, by measuring the response of the resistance to gate voltage in the absence of a magnetic field [29-31], and from quantum oscillation in magnetic fields [1, 32-34]. In this study, we studied the electronic band structure of high-quality samples of graphene consisting of one to seven layers by measuring quantum oscillations and found systematic variations in the band structure.

## Results
### Sample characterization

Figures 1 **b** and **c** show the schematic structures of the samples and an optical micrograph of a typical sample. Graphene was formed on a high quality $h$-BN flake [11, 12], which has an atomically flat surface, or encapsulated with $h$-BN flakes [13]. The number of layers and stacking were determined by examining the G' peak appearing at approximately 2700 cm$^{-1}$ in the Raman spectra [35], topography obtained using an atomic force microscope, and color intensity analysis of a digitized optical micrograph (see the Supplementary Information). The Raman G' peak for AB-stacked graphene exhibited a systematic variation with the numbers of layers [28, 35], as shown in Figure 1 **d**. We measured the Raman spectra for many graphene flakes. The shapes of the G' peaks were reproducible from sample to sample and consistent with the reports from other groups [35, 36], except for an offset in the background signal and slight offset in the Raman shift. This enabled us to accurately determine the number of layers and stacking. The carrier density was varied using a back gate built using a conducting Si substrate covered with its oxide [1]. Typical electrical mobilities ranged between 30,000 and 100,000 cm²/Vs.

### Landau fan diagram

Low-temperature magnetoresistance measurements showed Shubnikov-de Haas (S-dH) oscillations arising from Landau quantization. Figure 2 shows mappings of the derivative of magnetoresistance ($dR_{xx}/dB$) for samples with one to seven layers, as a function of gate voltage and magnetic field. These fan diagrams appear to be the finger-prints of graphene as to its specific layer number and stacking because they reflect specific band structures. In particular, the streaks and bright spots arise from the Landau levels and their crossings. They vary strikingly with the number of layers and directly reflect different electronic structures. Although the results for samples with one to four layers have been already reported in the literature [32-34, 37, 38], we will briefly review their features here to discuss the systematic variation of the electronic structure in multilayer graphene. As shown in Figure 2, mono- and bi-layer graphene have simple fan-shaped structures [1, 37, 38] because they have a set of Landau levels for an electron band and a hole band. The fan-shaped diagram for the monolayer originates from the fact that each Landau levels has a degeneracy of $4eB/h$. The monolayer and the bilayer are similar to each other but different in terms of the zero-mode Landau levels that appear at the charge neutrality point near $V_g \approx$ 0 V: the degeneracy for bilayer graphene is twice that for a monolayer [1], and it manifests itself as a difference in the widths of the zero-mode Landau levels. Trilayer graphene shows one fan-shaped structure arising from bilayer graphene, as well as a monolayer band whose Landau levels can be seen at low magnetic fields [32]. The Landau levels of the monolayer bands appearing in the fan diagram are quite different from those found in in monolayer graphene. Since monolayer and bilayer bands in trilayer graphene show different dispersion relations, the carrier density in the monolayer band is much less than that in the bilayer bands. Therefore, the Landau levels of the monolayer band reach the quantum limit at much lower magnetic fields compared to the bilayer band. For four or more layers, *i.e.*, $N \geq$ 4, Landau fan diagrams become increasingly complicated with increasing number of layers. As can be seen in the figure, bright spots and streaks near the charge neutrality point form complicated patterns, which differ layer to layer. This variation probably originates from the fact that that the Landau level structures are different for different layer numbers. Despite their complexity, we could find monolayer bands at odd layer numbers at low magnetic fields, as indicated by red marks in the figure. In the hole regime, (i.e., $n_{osc} < 0$) conspicuous streaks are observed at low magnetic fields. However, in the electron regime, one can discern streaks corresponding to monolayer bands only in trilayer graphene. This electron-hole asymmetry in the band structure arises possibly from the energy offset at the bottoms of the bands. These monolayer bands do not appear for even layers, *i.e.*, two, four and six-layer graphene.

## FFT results

The pattern that grows in complexity with increasing layer number in the vicinity of $n_{\text{tot}} = 0$ is possibly due to the formation of semi-metallic band structures because the Landau level crossings originate from electron and hole bands that are offset in energy. Here we will not go into the details of these structures, which will be reported elsewhere. Rather, we will discuss the general features of the band structure by taking the Fourier transform of the S-dH oscillation. The Landau level structure away from the charge neutrality point is rather simple, which enables us to obtain physical insights by estimating the number of bands and their dispersions. S-dH oscillations are periodic against the reciprocal of the magnetic fields and their period $\Delta(1/B)$ is relevant to the area $A$ of the energy contour (Fermi surface) of a band, from which the carrier density can be calculated using the formula, $n_{osc} = 4 \times A/(2\pi)^2 = 4 \ (e/h)/\Delta(1/B)$. We employed a fast Fourier transform (FFT) to observe the periodicity. Figure 3 shows the Fourier spectra as a function of the total carrier density $n_{\text{tot}}$ induced by the back-gate voltage. Here the total carrier density was calculated using calibrated gate capacitances, and the FFT frequency was scaled by a factor of $4 \ (e/h)$, which directly indicates the carrier density for low energy bands in graphene. In the figure, some noticeable spectra that reflect the carrier density of the Fermi surfaces appear at a given chemical potential tuned by the gate voltage. Other peaks can be assigned to higher-order harmonics whose frequencies can be expressed as multiples of the base frequency or their sums (see the Supplementary Information). These samples show a characteristic layer number dependence. It is clear that FFT spectra for samples with $2N$ and $2N + 1$ ($N = 1, 2, 3$) layers are somewhat similar to each other in terms of the $n_{tot}$ dependence of $n_{osc}$ except for the presence of bands with a small carrier density in odd layers. This should occur because $2N$ and $2N + 1$ layer graphene have $N$ bilayer bands: for example, two and three layer each have one bilayer band, four and five layer graphene each have two bilayer bands, *etc.* Reflecting dispersion relations, the density of states of the monolayer band is much less than that of bilayer bands. The relative intensity of the S-dH oscillation arising from the monolayer band decreased as the layer number increased. This is because the carrier density of the monolayer band is reduced because the number of bilayer bands is increased.

In addition to this even-odd layer number effect relevant to the monolayer band, there are small differences in the $n_{osc}$ of bilayer bands between the samples with $2N$ and $2N + 1$ ($N = 1, 2, 3$) layers. This difference partly stems from the dependence of the band mass on the layer number; although the number of bilayer bands in $2N$ layers equals that of $2N + 1$ layers, values of the mass in those bands are expected to be slightly different [22]. Moreover the difference can also result from the fact that the energy offset value of each band is somewhat dependent on the layer number. [20]

Discussion

### Theoretical calculation and screening of gate potential

We performed a calculation based on the tight-binding model. A simple calculation of the dispersion relation using all the Slonczewski-Weiss-McClure (SWMcC) parameters of graphite could approximately explain the experimental result obtained for $N < 4$. However for $N \geq 4$, the disagreement became progressively larger as the layer number increased (see the Supplementary Information). Fine tuning of the SWMcC parameters did not significantly change the situation. Much better agreement between experiment and calculation was obtained when the potential due to screening by gate induced carriers was considered. Figure 4 shows the results. In our experiment, the carrier density was tuned by the external gate voltage. This is basically equivalent to the behavior of a capacitor formed by the graphene, insulator (*h*-BN), and gate electrode. A simple capacitor model tells us that the induced charge is located at the surface of the metals and that there are no electric fields inside the gate electrode. However, in the case of an atomic layer material, one must take screening of the induced charge into account, as was discussed in Refs. [27, 39-41] Hence in the calculation, we assumed screening of the form $\propto \exp(-d/\lambda)$ where $\lambda$ is the screening length and $d$ is distance from the surface. The parameter $\lambda$ was adjusted so that calculations fit the experiments (see the Supplementary Information). The effect of screening becomes increasingly evident as the layer number increases. The screening lengths are equivalent to approximately one to two layers, which approximately agrees with the theoretically predicted value [41].

Since the beginning of graphene research, there have been investigation into screening of gate induced carriers [27, 40, 41], and the short screening lengths of a few layers have been regarded as having an effect on the electronic property of multilayer graphene [42]. Since the screening length has been expected to be equivalent to approximately a few layers in multilayer graphene, it was assumed that only a few layers could be measured if one used gate electrodes to vary the carrier density of the samples [42]. However, our observations show an even-odd layer number effect and number-of-layers-dependent fan diagrams, which reflects the particular electronic band structures of graphene consisting of four to seven layers. We speculate that these findings indicate that, for at least up to seven layers, coherent quantum states associated with the band structure form along the layer direction, *i.e.*, perpendicular to the sample plane, even in the presence of a gate induced potential variation.

The present results answer the question of how graphene becomes graphite as the layer numbers increases. Generalizing this argument, we can conclude that monolayer bands exist even in graphene with larger numbers of odd layers, and presumably even in graphite as long as AB-stacking is preserved in the entire sample and that the number of layers is odd. However, finding a monolayer band would become increasingly difficult because of the large number of bilayer bands.

### On other stackings and other systems

We have limited our concern to band structures in AB-stacked graphene. For a layer number larger than 2, ABC-stacking, which is not equivalent to ABA-stacking, is possible and is expected to show different band structures. The number of possible stacking increases with the number of layers. Each stacking is expected to show different band structures [24, 43]. Some of these band structures with small layer numbers have been identified from optical measurements. However, transport experiments have only been carried out for ABC-stacked trilayer graphene [31, 44].

Further studies of graphene with various layer stackings will contribute to a comprehensive understanding of the transition from atomic layers to the bulk.

A layer number dependence in the electronic band structures was also reported in metal dichalcogenides: the energies of the bottoms of the band were found to vary systematically with the number of layers [45-47]. Moreover, a similar but different even-even odd layer number effect was recently observed in $MoS_2$ atomic layer devices [48]; the reported Landau levels showed a valley Zeeman effect in samples with odd layer numbers, and a spin Zeeman effect in samples with even layer numbers. Another example is the topological insulator $Bi_2Se_3$. Topologically protected surface states, which appear in the bulk, showed mixing as the number of layers decreases [49]. Hence, more interesting phenomena might exist that relate atomic layers and bulk materials.

## Summary

Shubnikov-de Haas oscillations in AB-stacked graphene with one to seven layers were reported. Graphene exhibited characteristic Landau level structures in the fan diagram, which reflected the dependence of the electronic states on the number of layers. FFT analysis of the periodicity in the $1/B$-dependence of the magnetoresistance revealed a systematic change in the band structures; in particular, an even-odd layer number effect relevant to the electronic band structure was observed, where a linear band was observed only in odd-layer graphene. The band structure became more sensitive to the gate electric field as the layer number increased.

## Methods

Graphene samples are mechanically exfoliated from Kish graphite. The graphene is then transferred onto a thin h-BN flake, which was prepared by mechanically exfoliating a high quality *h*-BN crystal. Some samples are encapsulated with *h*-BN flakes. Graphene samples were patterned by using electron beam lithography. Electric resistance measurements were carried out using a standard lock-in technique. Magnetic fields were applied by using a superconducting solenoid.

## Acknowledgements

This work was supported by MEXT KAKENHI Grant Number JP25107003.

## Contributions

RY conceived the experiments. KW and TT made the high-quality h-BN crystals. HT, RE, TN and ST made the graphene samples. RY, HT, RE, TN and ST carried out low-temperature experiments. RY analyzed the data and calculated the band structure. RY wrote the manuscript.

## Competing Interests

The authors declare no conflicts of interest associated with this manuscript.

# Figure Legends

**Fig. 1**   Sample characterization. |
**a.** Schematic diagram of the band structure of AB-stacked graphene. **b** Sample structure. Some graphene samples were encapsulated with *h*-BN flakes. Other samples were not encapsulated but transferred onto the top of a thin *h*-BN flake. **c.** Optical micrograph of a typical encapsulated sample. **d.** (Top) Example of Raman spectra. (Bottom) Variation in Raman G' band spectra in AB-stacked graphene with increasing layer number from 3 to 8. Data are scaled and offset to better compare the shapes of the peaks.

**Fig. 2**   Landau fan diagram. |
  Landau fan diagram for AB-stacked graphene. Measurements were carried out at $T$ = 4.2 K.   The derivative of the transverse resistivity with respect to the magnetic field is plotted on a color scale as a function of the total carrier density which was tuned by the gate voltage. To improve visibility the cube root of $\frac{dR_{xx}}{dB}$ is plotted. Red marks indicate the positions of the peaks arising from monolayer bands.

**Fig. 3**   Carrier density of each band. |
  FFT spectra of magnetoresistance plotted against total carrier density $n_{tot}$. The magnitude of the spectra is plotted with color. The frequency of the FFT (vertical axis) has been converted to have dimensions of carrier density.

**Fig. 4** Comparison with band calculation. |
  The dependence of $n_{\text{osc}}$ on total carrier density $n_{\text{tot}}$ for each band was calculated for graphene with three to seven layers using a Hamiltonian based on the effective mass approximation. The results are displayed as symbols, while the lines are guides for the eye. Carrier density was calculated from the energy contour of the dispersion relation at zero magnetic field.   The SWMcC parameters for this calculation were $\gamma_0 = 3.$ eV, $\gamma_1 = 0.45$ eV, $\gamma_2 = -0.023$ eV, $\gamma_3 = 0.3$ eV, $\gamma_4 = 0.04$ eV, $\gamma_5 = 0.04$ eV, and $\Delta' = (\Delta - \gamma_2 + \gamma_5) = 0.032$eV. Screening length λ was 0.33, 0.33, 0.33, 0.43, and 0.35 nm for layers three to seven, respectively. The SWMcC parameters for this calculation better reproduced the experimental results compared to the parameters used for graphite. The peaks for the bilayer are labeled as $fb_1$, $fb_2$, and $fb_3$. The peak of the monolayer band is labeled fm.

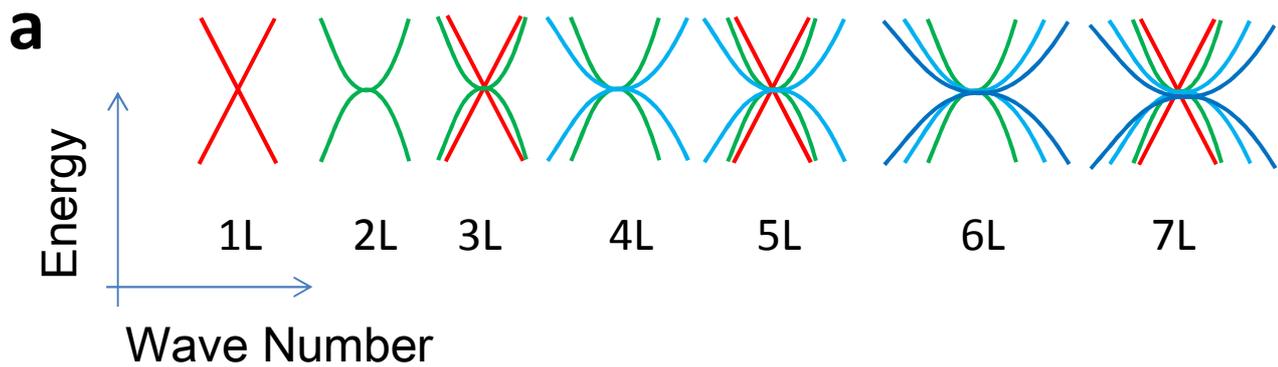
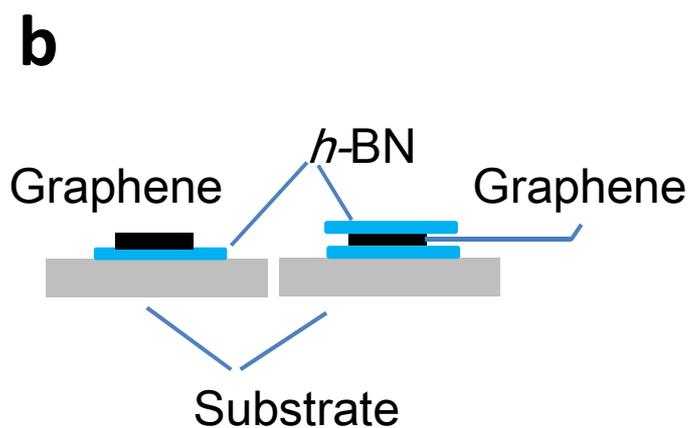
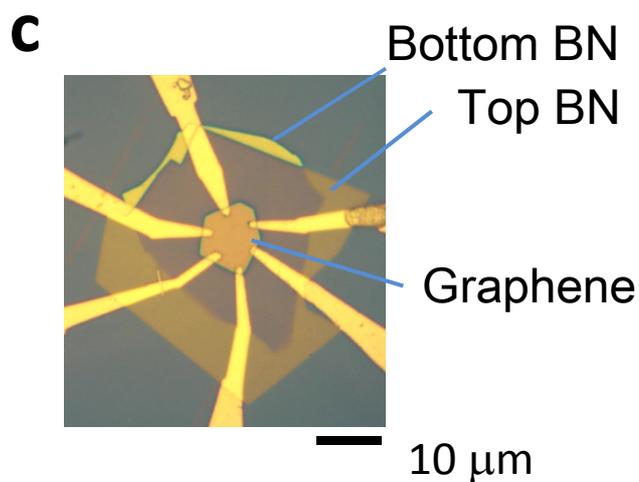
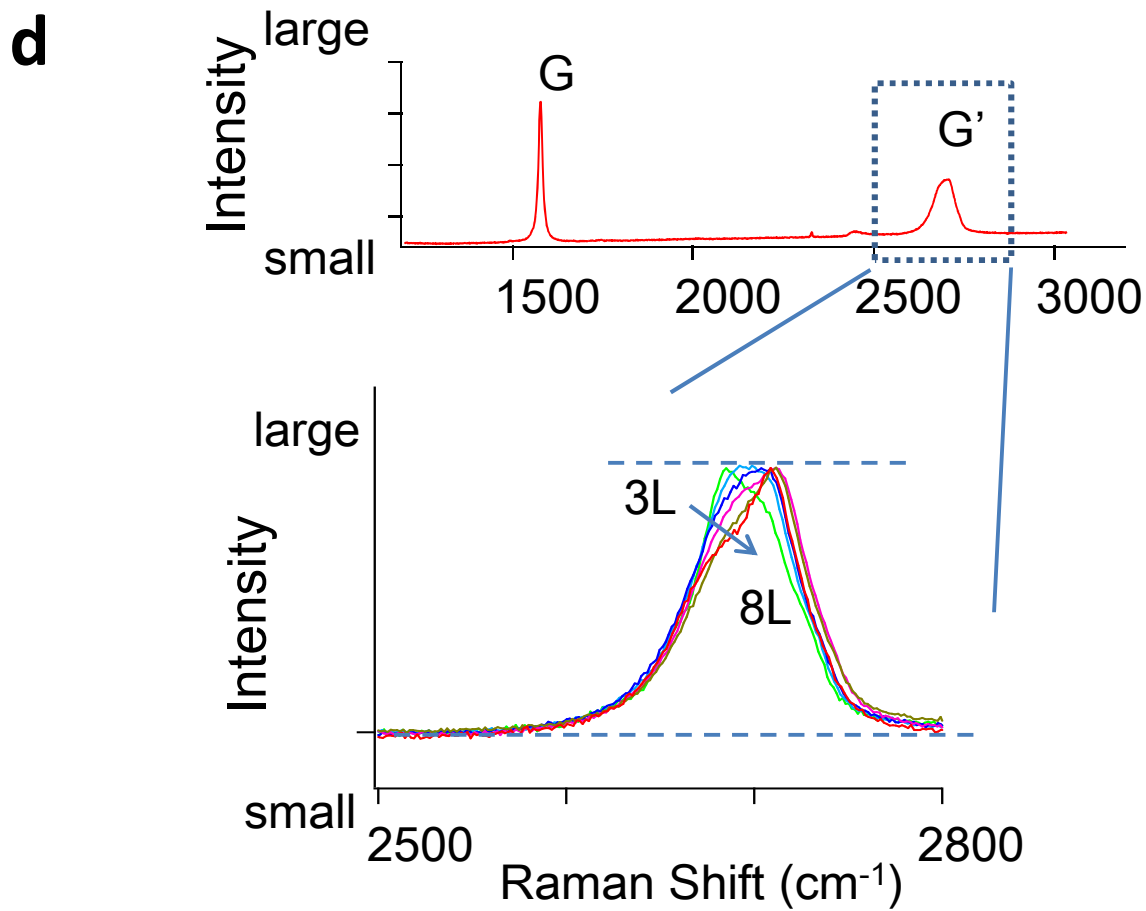

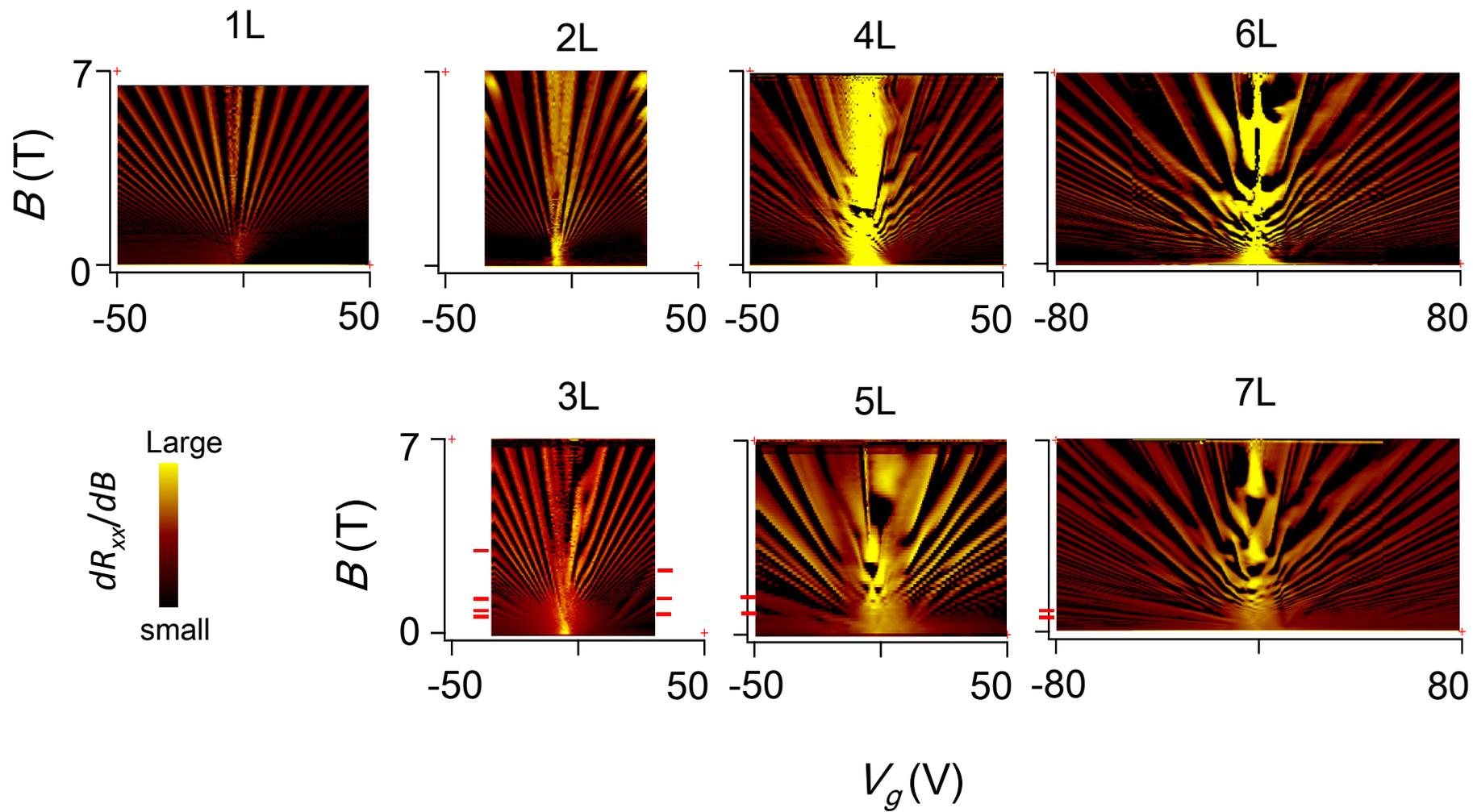

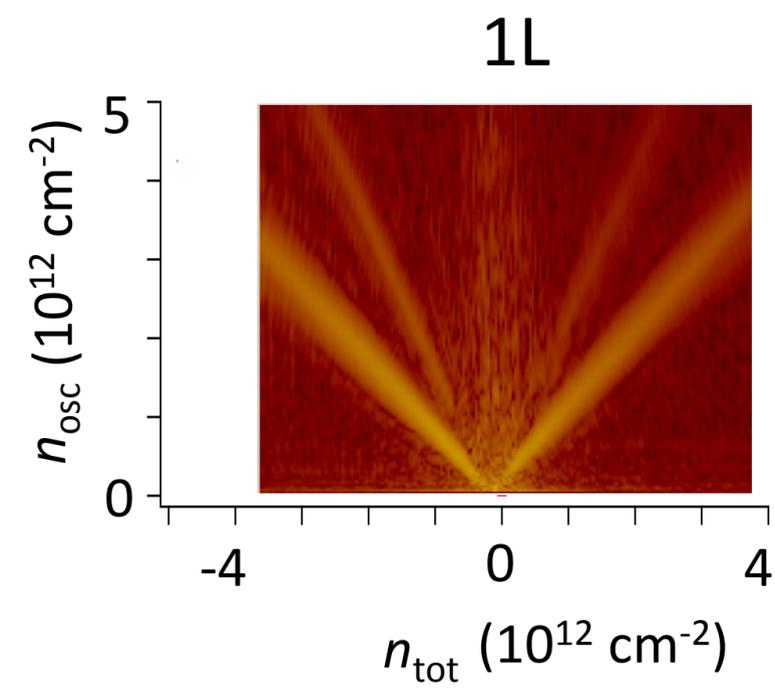
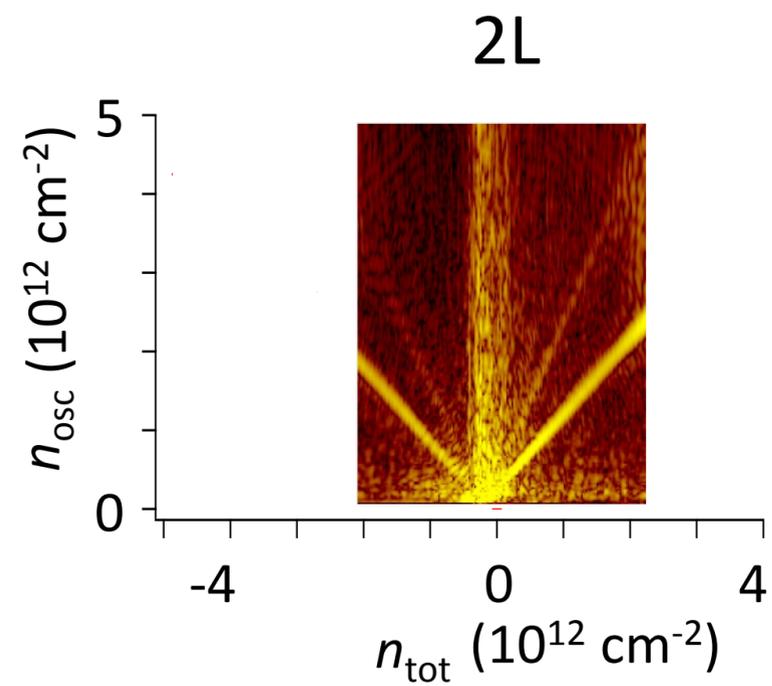
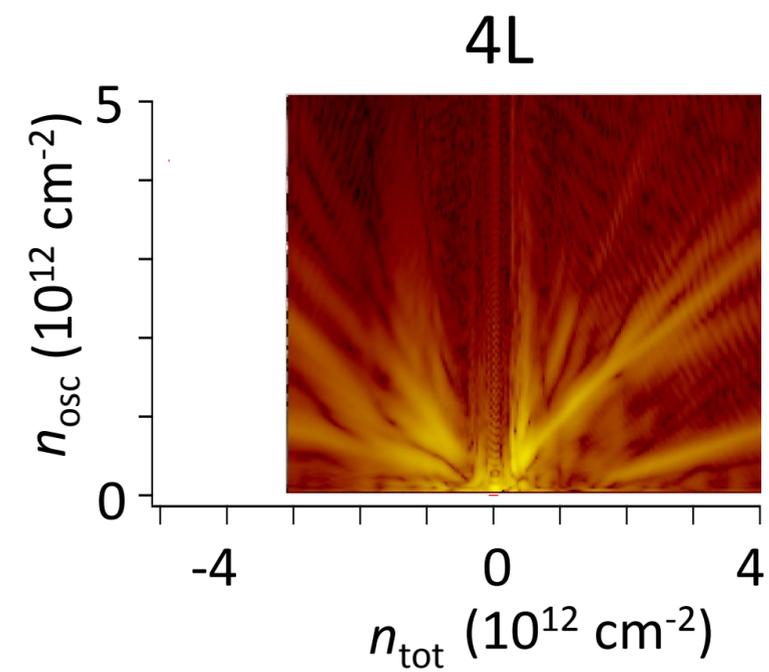
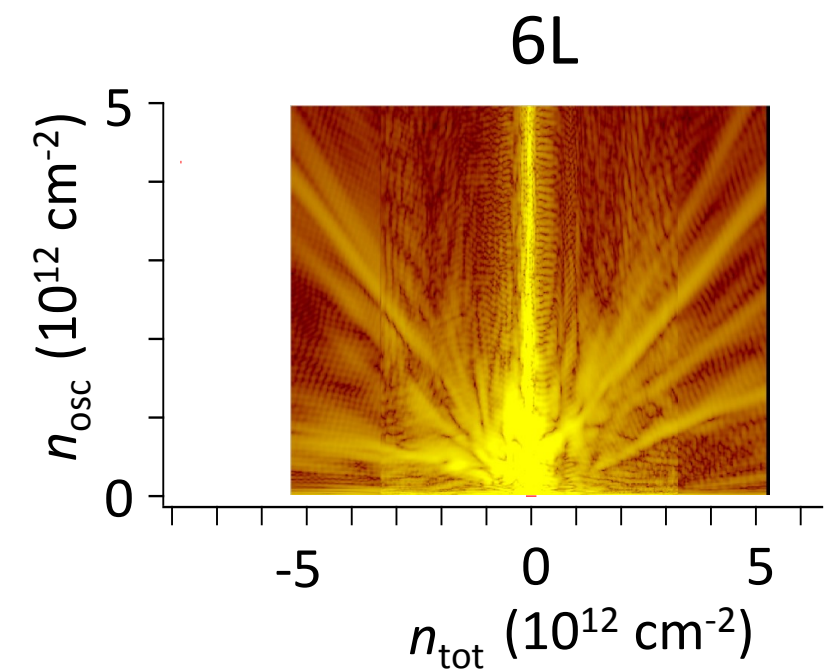
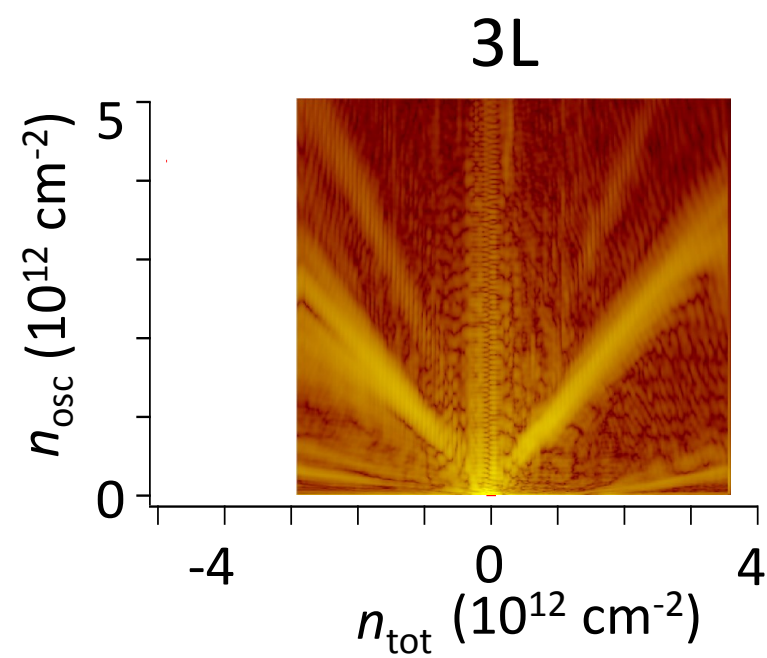
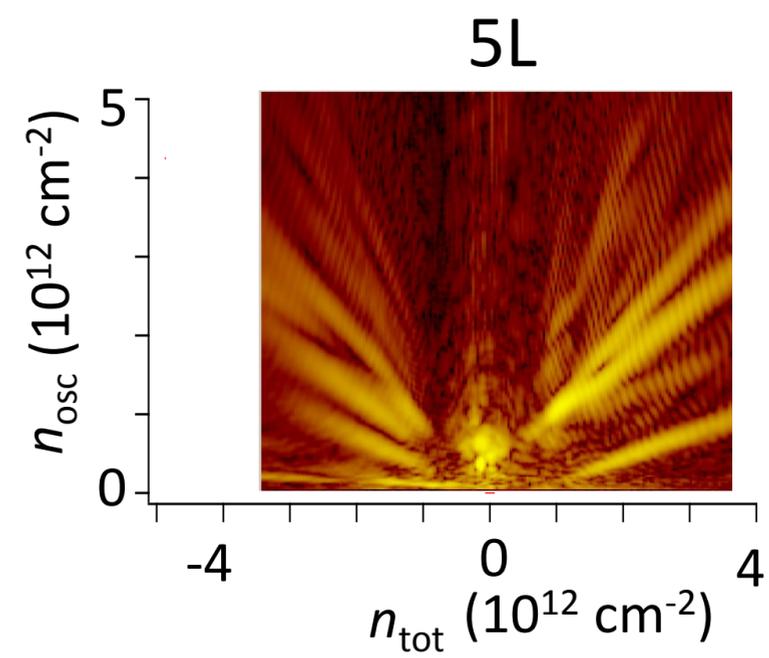
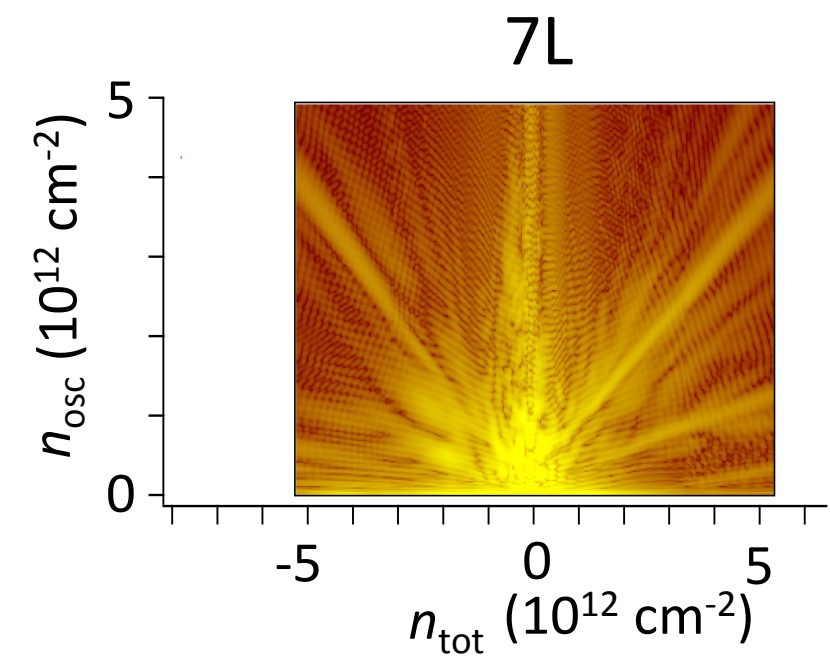

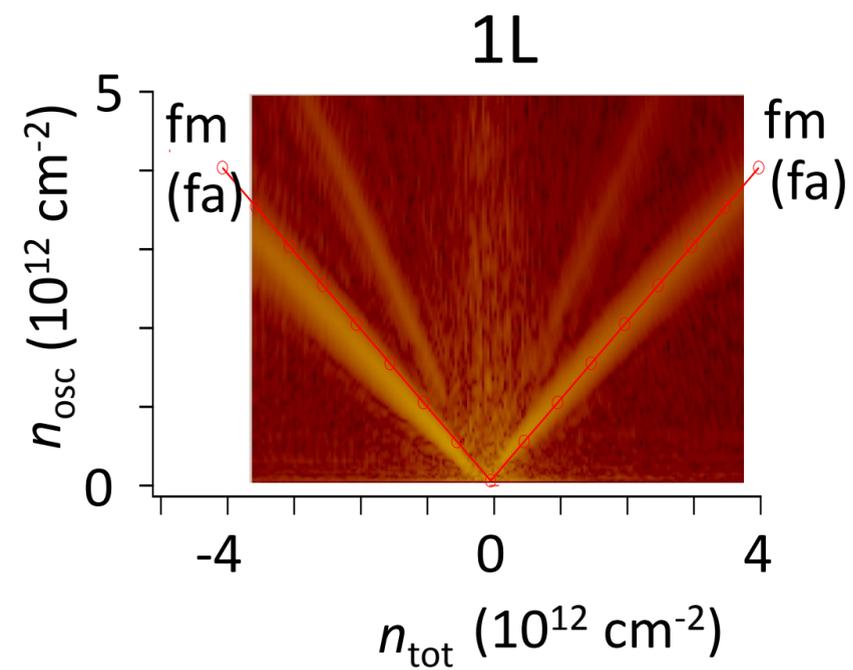
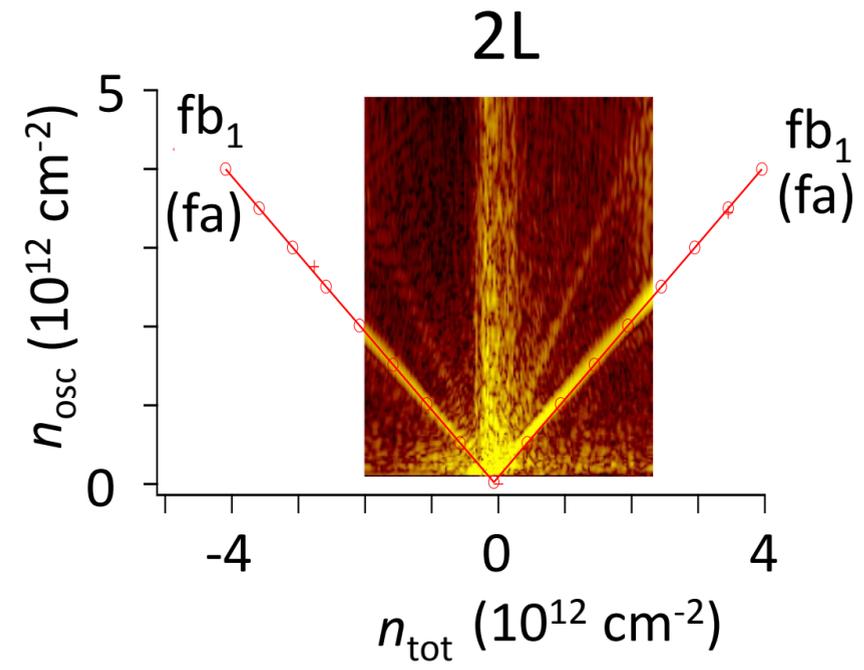
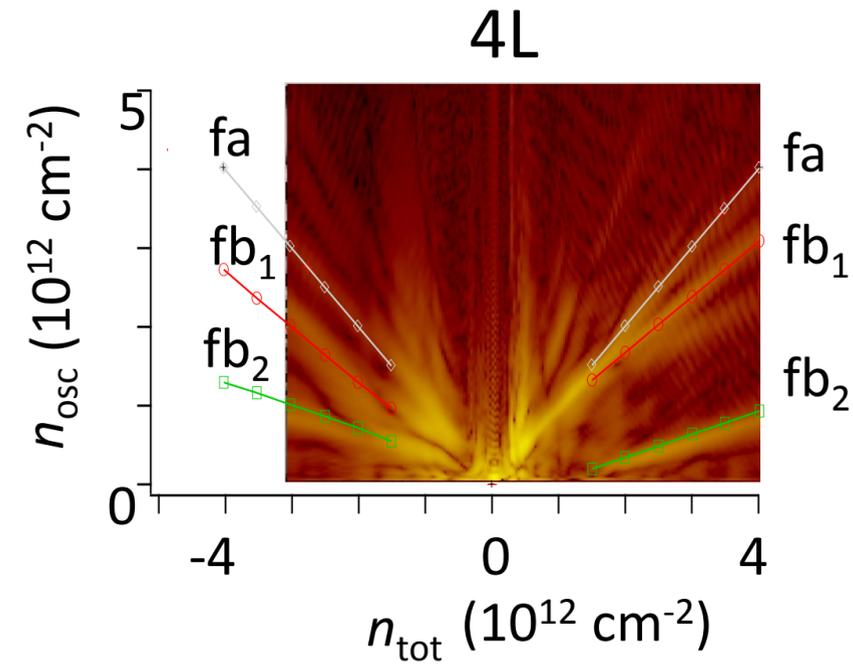
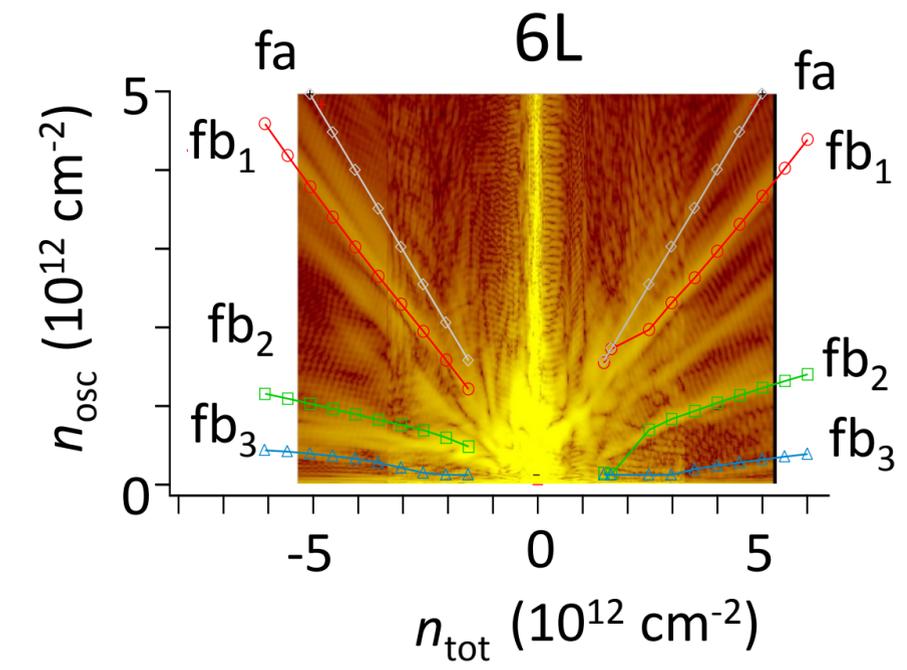
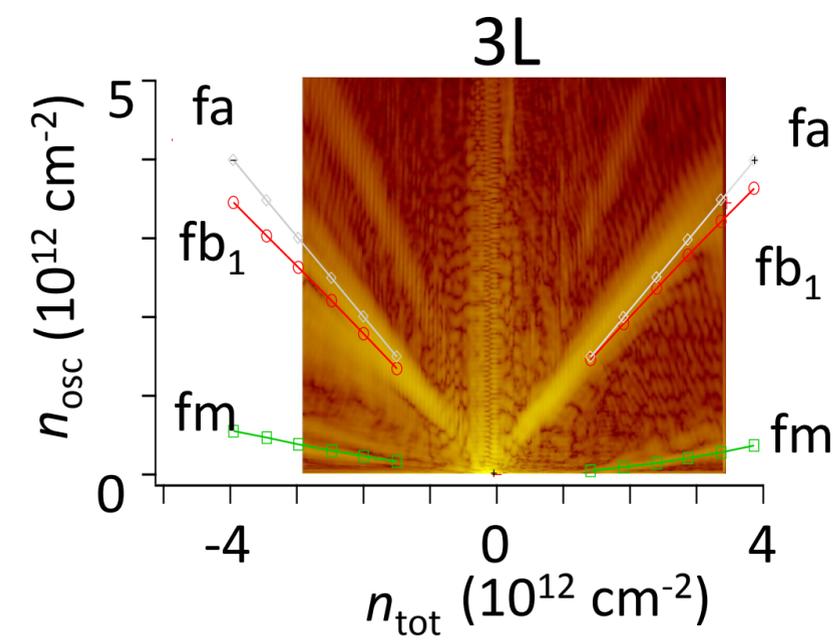
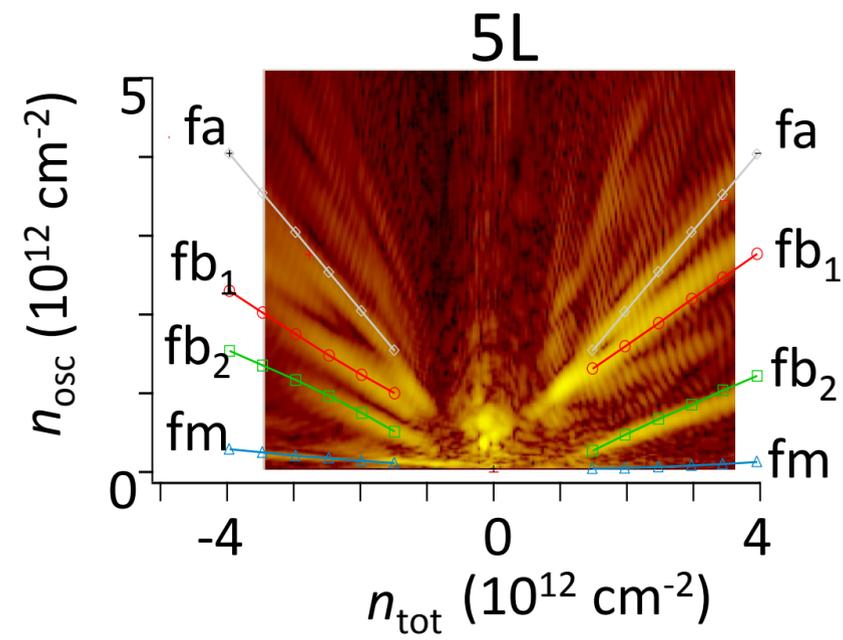
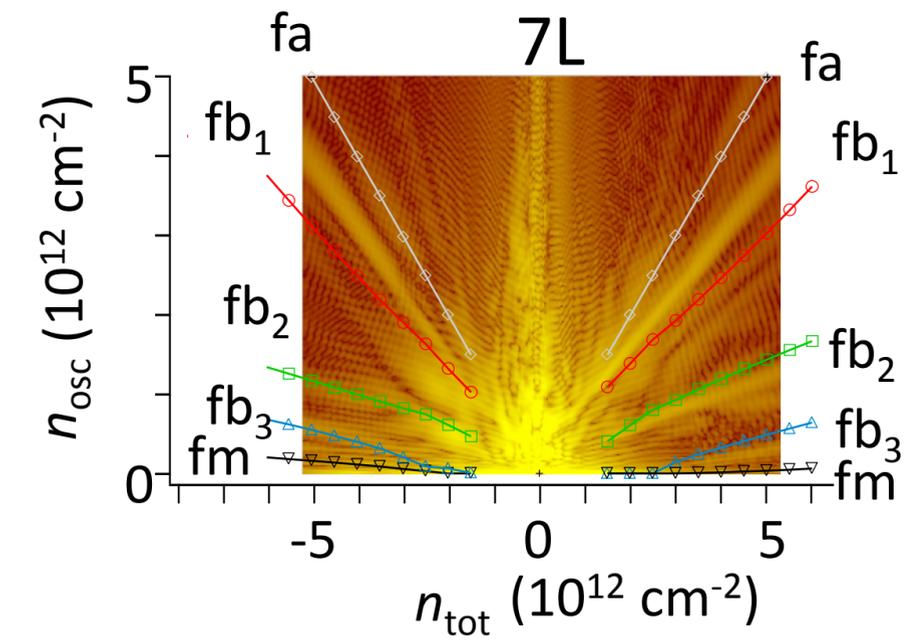